\documentclass[runningheads, 10pt]{llncs}
\usepackage[T1]{fontenc}
\usepackage{graphicx}
\usepackage{amsmath,amsfonts}
\usepackage{algorithmic}
\usepackage{thmtools,thm-restate}
\usepackage{enumitem}
\usepackage{amssymb}
\usepackage{graphicx}
\usepackage{comment}
\usepackage{cite}
\usepackage{textcomp}
\usepackage{xcolor}
\usepackage{wrapfig}
\usepackage{xspace}
\usepackage{multirow}
\usepackage{tabularx}
\usepackage[thinlines]{easytable}
\usepackage{listings}
\usepackage{hyperref}
\usepackage{siunitx}
\usepackage[nonumberlist,acronyms,nogroupskip]{glossaries}

\hypersetup{
    colorlinks,
    linkcolor={red!50!black},
    citecolor={blue!50!black},
    urlcolor={blue!80!black}
}

\makeatletter
\let\oldmakefirstuc\makefirstuc
\renewcommand*{\makefirstuc}[1]{%
  \def\gls@add@space{}%
  \mfu@capitalisewords#1 \@nil\mfu@endcap
}
\def\mfu@capitalisewords#1 #2\mfu@endcap{%
  \def\mfu@cap@first{#1}%
  \def\mfu@cap@second{#2}%
  \gls@add@space
  \oldmakefirstuc{#1}%
  \def\gls@add@space{ }%
  \ifx\mfu@cap@second\@nnil
    \let\next@mfu@cap\mfu@noop
  \else
    \let\next@mfu@cap\mfu@capitalisewords
  \fi
  \next@mfu@cap#2\mfu@endcap
}
\makeatother

\newcommand\blfootnote[1]{%
  \begingroup
  \renewcommand\thefootnote{}\footnote{#1}%
  \addtocounter{footnote}{-1}%
  \endgroup
}

\glsdisablehyper

\definecolor{light-gray}{rgb}{1,0,0} 
\newcommand{\ulg}{\textcolor{light-gray}{$\blacktriangle$}\xspace} 

\newacronym{SotA}{SotA}{state-of-the-art}
\newacronym{HE}{HE}{homomorphic encryption}
\newacronym{MPC}{MPC}{multi-party computations}
\newacronym{PPML}{PPML}{privacy-preserving machine learning}
\newacronym{MSE}{MSE}{mean square error}
\newacronym{ML}{ML}{machine learning}
\newacronym{CNN}{CNN}{convolutional
neural network}
\newacronym{MLP}{MLP}{multi-layer perceptron}
\newacronym{NN}{NN}{neural network}
\newacronym{SIMD}{SIMD}{single-instruction, multiple-data}
\newacronym{FC}{FC}{fully connected}
\newacronym{GC}{GC}{garbled circuit}
\newacronym{OT}{OT}{oblivious transfer}
\newacronym{SS}{SS}{secret sharing}
\newacronym{NAS}{NAS}{neural architecture search}
\newacronym{RMSE}{RMSE}{root-mean-square error}

\newcommand{\viz}{{namely,}\xspace}
\newcommand{\vs}{{vs.}\xspace}
\newcommand{\etal}{{et al.}\xspace}

\newcommand{\eg}{{e.g.,}\xspace}
\newcommand{\ie}{{i.e.,}\xspace}

\newcommand{\mysec}[1]{\vspace{3pt}\noindent\textbf{#1.}\xspace}

\newcommand{\meth}[1]{{\textbf{\underline{#1}}}}

\newcommand{\hl}[1]{\textcolor{red}{#1}}
\renewcommand{\hl}[1]{#1}

\newcommand{\FRAME}{HE-PEx\xspace}

\begin{document}
\title{Efficient Pruning for Machine Learning under {H}omomorphic {E}ncryption\vspace{-15pt}} 
%
%
\author{Ehud Aharoni\inst{1}, Moran Baruch\inst{1}, Pradip Bose\inst{2}, Alper Buyuktosunoglu\inst{2},\\Nir Drucker\inst{1}, Subhankar Pal\inst{2}, Tomer Pelleg\inst{1}, Kanthi Sarpatwar\inst{2},\\Hayim Shaul\inst{1}, Omri Soceanu\inst{1}, and Roman Vaculin\inst{2}\vspace{-2.5pt}}
\authorrunning{E.\@ Aharoni et al.}
%

\institute{IBM Research -- Israel\\\and
IBM T.\@ J.\@ Watson Research Center, USA\\
\vspace{0pt}\email{subhankar.pal@ibm.com}, \email{nir.drucker@ibm.com}\vspace{-10pt}}

\maketitle              
\blfootnote{The Version of Record of this contribution is published in ESORICS 2023, and is available online at \url{https://doi.org/10.1007/978-3-031-51482-1_11}.}
\begin{abstract}\vspace{-2.5pt}
Privacy-preserving machine learning (PPML) solutions are gaining widespread popularity.
Among these, many rely on homomorphic encryption (HE) that offers confidentiality of the model and the data, but at the cost of large latency and memory requirements.
Pruning neural network (NN) parameters improves latency and memory in plaintext ML but has little impact if directly applied to HE-based PPML.

We introduce a framework called \textbf{HE-PEx} that comprises new pruning methods, on top of a packing technique called tile tensors, for reducing the latency and memory of PPML inference. 
HE-PEx uses \textit{permutations} to prune additional ciphertexts, and \textit{expansion} to recover inference  loss.
We demonstrate the effectiveness of our methods for pruning fully-connected and convolutional layers in NNs on PPML tasks, \viz image compression, denoising, and classification, with autoencoders, \acrfullpl{MLP} and \acrfullpl{CNN}.

We implement and deploy our networks atop a framework called HElayers, which shows a \textbf{10--35\%} improvement in inference speed and a \textbf{17--35\%} decrease in memory requirement over the unpruned network, corresponding to \textbf{33--65\%} fewer ciphertexts, within a \textbf{2.5}\% degradation in inference accuracy over the unpruned network.
Compared to the state-of-the-art pruning technique for PPML, our techniques generate networks with \textbf{70}\% fewer ciphertexts, on average, for the same degradation limit.

\end{abstract}
\section{Introduction}\label{sec:intro}

Data privacy and confidentiality are crucial in today's information-driven world.
Outsourcing sensitive data to a third-party cloud environment, while complying with regulations such as GDPR \cite{GDPR}, is the need of the hour for many companies and organizations, such as banks and medical establishments. 
One promising solution is the use of \gls{HE}, which allows for the evaluation of certain functions on encrypted inputs.
Corporations and academic organizations are already  investing resources into developing secure and efficient solutions~\cite{hebench,standard}. 
However, the major challenge constraining faster adoption is that HE applications involving a large number of operations are significantly slower and have higher memory requirements than their plaintext counterparts~\cite{complex}. 

{Non-client--aided}, or non-interactive, \gls{HE} is a popular choice for \gls{PPML} inference and runs on systems that involve at least one client and one server.
Here, the clients desire the privacy of their input data (\eg images), and the server performs inference over an encrypted \gls{ML} model, which has been trained with proprietary data. 
Typical HE schemes perform computation using a \gls{SIMD} paradigm, for speed and efficiency. For instance, in CKKS~\cite{CKKS2017}, $\frac{N}{2}$ input elements are {packed} into a polynomial of degree $N$. 
A recent work, called HElayers~\cite{helayers}, proposes efficient packing for multi-dimensional inputs using a technique called \textit{tile tensors}. 
This method decomposes the input into {tiles} that are then each packed into a separate ciphertext.

In order to improve inference speed and memory, \glspl{NN} use a common technique called \textit{pruning}, for convolutional filters, \gls{FC} weights, or even entire nodes in a trained model~\cite{pruning2,pruning3}.
The challenge arises when we attempt to re-use plaintext \gls{ML} pruning strategies for non-client--aided \gls{HE} under \gls{SIMD} packing.
Na\"ive application of such a strategy fails due to the fundamental reason that while pruning may introduce zeros, if the zeros lie in a ciphertext with even one other non-zero, then the ciphertext cannot be eliminated.
Thus, pruning 85\% of the \gls{NN} weights (85\% sparsity), for instance, may lead to the elimination of only 17\% of the ciphertexts (17\% \textit{tile sparsity}).

\mysec{Our Solution}
We propose a framework called \textbf{\FRAME} for pruning homomorphically encrypted \glspl{NN} that produces deployment-ready models with superior tile sparsity and small inference losses.
Our contributions are as follows:
\begin{itemize}[noitemsep,topsep=0pt]
    \item 
    We introduce a  set of novel methods to perform pruning for \gls{NN} models under \gls{HE} with \gls{SIMD} packing, which combines four main primitives: \textit{prune}, \textit{permute}, \textit{pack}, and \textit{expand}. We propose a novel co-permutation algorithm that reorders the structure of the network to improve the tile sparsity, without impacting the network accuracy and with minimum overhead.

    \item We integrate our techniques into HElayers~\cite{helayers} to develop a holistic framework that takes in an \gls{NN} model and automatically produces a pruned and packed network that meets various user constraints, such as accuracy, latency, throughput, and memory requirement.
    
    \item We adapt a \gls{SotA} pruning method, Hunter~\cite{hunter}, to support \glspl{NN} under non-interactive \gls{HE}. 
    This serves as a baseline for our experiments.
    
    \item
    We implement and compare several schemes on four PPML networks with three datasets for different tile shapes, comparing inference accuracy and ciphertext reduction fractions. 
    We additionally report our results on a large \gls{NN} trained to detect COVID-19 from CT scans under HE.
\end{itemize}
Our techniques produce networks with tile sparsities of up to \textbf{95\%} (average \textbf{61\%}) across the datasets and \glspl{NN}, within a limit of 2.5\% degradation in network accuracy/loss.
These improve upon the \gls{SotA} by an average of \textbf{70}\%.
By leveraging this sparsity, we demonstrate a \textbf{10--35\%} (\textbf{17--35\%}) improvement in measured inference speed (resp., memory), compared to the unpruned model, for a privacy-preserving image denoising application run using HElayers.
For a higher degradation limit of 5\%, these improvements are up to \textbf{41\%} (\textbf{41\%}).


\section{Related Work}\label{sec:related}

Yang~\etal~\cite{fpgaprune} propose a new FPGA design that can efficiently handle sparse ciphertexts.
It assumes a network that is pruned based on some standard pruning method and leverages the fact that some ciphertexts happen to encrypt zeros. 
Chou \etal~\cite{cryptonets-pruning} use pruning to reduce the number of weights, and quantization to convert them to powers-of-two, resulting in sparse polynomial representations.
However, their method applies to encoded plaintexts only and is demonstrated for PPML use-cases where the model is unencrypted.
Popcorn~\cite{pailler-pruning} considers the pruning of convolution layers in the context of Paillier encryption~\cite{paillier} and XONN~\cite{xonn} considers it in the context of \glspl{GC}, however their algorithms are different from ours and do not consider packing with tile tensors. 
\gls{NN} pruning was also considered by Gong \etal~\cite{pp-pruning1} in the context of \gls{PPML}, where the goal was to maintain the privacy of the data in the training dataset. However, neither the model nor the data is encrypted and their privacy aspect lies with the data used for training, not inference. 

A recent work called Hunter~\cite{hunter} applies pruning to \glspl{NN} in an \gls{HE}-packing--aware manner.
It considers the way the data is encoded and encrypted into the \gls{HE} ciphertexts, which  reduces the number of ciphertexts during computation.
The packing-based pruning method of Hunter uses the specific packing choice of the client-aided solution GAZELLE~\cite{GAZELLE2018}, where the client evaluates the activation functions. 
\FRAME, instead, targets a non-interactive scenario in which the server computes the activation layers and, therefore, should perform an online sequence of encrypted matrix multiplications.
We adapt the Hunter scheme to this scenario, naming it P2T, and compare with our methods.
While we focus on non-interactive scenarios, we argue that \FRAME is equally valuable in a  client-aided case.
The main differences between \FRAME and Hunter are:
\begin{itemize}[noitemsep,topsep=0pt]
    \item \FRAME targets a non-client-aided solution using HElayers~\cite{helayers}, whereas Hunter targets a client-aided solution called GAZELLE~\cite{GAZELLE2018}.
    \item \FRAME introduces a novel permutation algorithm to improve tile sparsities.
    \item \FRAME uses an expansion technique that improves the inference accuracy of the \gls{NN} models without affecting their tile sparsities.
    \item \FRAME combines the prune, permute and expand strategies into a holistic framework that yields better results than simple packing-aware pruning. 
\end{itemize}


\section{Background}\label{sec:bg}

This section provides background on \gls{HE} and a recent method of packing ciphertexts called tile tensors, which we leverage for efficient pruning in PPML inference. We conclude this section with our assumed threat model.

\subsection{Homomorphic Encryption}\label{sec:he}

An \gls{HE} scheme is an encryption scheme that enables computation on encrypted data. 
Modern \gls{HE} instantiations such as CKKS~\cite{CKKS2017} rely on the hardness of the Ring-LWE problem and support \gls{SIMD} operations. 
They provide the standard public-key encryption methods ($Gen$, $Enc$, $Dec$), where $Enc$ encrypts an $s$-slot vector $M$ to a ciphertext $[M]$ and $Dec$ decrypts a ciphertext $[M]$ to an $s$-slot vector. 
An \gls{HE} scheme is correct if for every vector $M$, $M = Dec([M])$ and is approximately correct (as in CKKS) if for some small $\epsilon > 0$ that is determined by the key it follows that $|M(i) - Dec([m])(i)| \le \epsilon$. 
In addition, CKKS provides the homomorphic addition ($\oplus$) and multiplication ($\cdot$) functions, where $Dec([M] \oplus [M'])_i \approx M_i + M'_i$, and $Dec([M] \cdot [M'])_i \approx M_i * M'_i$, respectively. 
It also provides a rotation ($Rot$) function, where $Dec(Rot([M],n))_i \approx M_{i+n \pmod s}$.

\mysec{Non-Interactive Homomorphic Encryption}
Many frameworks offer \gls{PPML} inference solutions using \gls{HE} or a combination of it with \gls{MPC} protocols~\cite{baruch2023sensitive}.
Protocols that exclusively use \gls{HE}, such as~\cite{helayers, hemet}, are called non-interactive or non-client--aided protocols. 

In the client-aided approach, the client assists the server, \eg to compute a non-polynomial function, such as ReLU, that is not supported natively in \gls{HE}. 
Here, the server asks the client to decrypt the intermediate ciphertext, perform the ReLU computation, and re-encrypt the data. 
This approach is implemented, for example, in GAZELLE~\cite{GAZELLE2018} and nGraph-HE~\cite{HET}. 
In client-aided solutions, the server utilizes \gls{MPC} to hide the intermediate results from the client.

The main drawbacks of client-aided solutions are that the client must stay online during the computation, and that the repeated data transfers between the client and the server may induce variability in inference latency. 
Moreover, prior works demonstrate that this approach may involve security risks~\cite{AkaviaVald21} and is susceptible to model-extraction attacks~\cite{muse}. 
To this end, our focus is on non-client-aided solutions, where the computation is done entirely under \gls{HE}.
However, we stress that our proposed methods are also applicable to interactive solutions. 

\subsection{Ciphertext Packing using Tile Tensors}\label{sec:tiletensors}

\gls{HE} schemes that operate on ciphertexts in a \gls{SIMD} fashion, such as CKKS~\cite{CKKS2017}, allow encrypting a fixed-size vector into a single ciphertext, and the \gls{HE} operations on the ciphertext are performed slot-wise on the elements of the plaintext vector.
The use of \gls{SIMD} also makes it faster to execute on modern processors that support efficient vector operations.
To leverage this, we pack several input elements in each ciphertext. 
The choice of packing method can dramatically affect the {latency} (computation time), {throughput} (number of computations performed per unit time), communication costs (\eg server-client bandwidth requirement), and memory requirement.

\setlength{\intextsep}{1pt}
\begin{wrapfigure}{r}{0.5\textwidth}\vspace{-12pt}
  \begin{center}
    \includegraphics[width=0.495\textwidth]{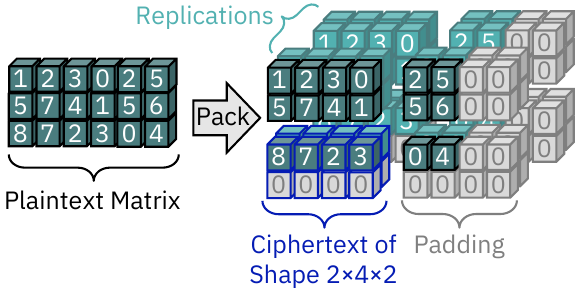}
  \end{center}
  \vspace{-15pt}
  \caption{An example of packing a $3$$\times$$6$ matrix in a tile tensor format that operates over 8 ciphertexts ($16$ slots each).
    The matrix is zero-padded over dimensions 1 and 2, and is replicated three times over dimension 3. See~\cite{helayers} for further information.}
  \label{fig:tiletensors}
\end{wrapfigure}

A recent work, HElayers~\cite{helayers}, proposes a mechanism called {\em tile tensor} that packs tensors (\eg matrices) into fixed-size chunks, called {tiles}.
The authors of~\cite{helayers} demonstrate the use of tile tensors to implement various \gls{NN} layers.
For example, they implement encrypted inference on an HE-friendly variant of AlexNet~\cite{alexnet}.
We target HElayers and the tile tensor method of packing due to its generality and flexibility, which allows running networks of diverse sizes under \gls{HE}.

HElayers demonstrates that the same tensor can be packed into tiles of different shapes, as long as they have the same size. 
For instance, while a matrix may be na\"ively packed into column-vectors or row-vectors, it can also be packed into 2D tiles, as long as the tile size matches the number of slots in the ciphertext.
In addition, tile tensors allow for other manipulations, such as duplicating elements along one or more dimensions.
This is necessary, for instance, when there is a batch dimension to the input and the weights need to be replicated to construct 3D tiles.
Fig.~\ref{fig:tiletensors} shows one way to pack a 3$\times$6 matrix into a 3D tile tensor with three replications of the original matrix.
Further, note that the security level of the scheme depends on the size of the tile and not on the individual dimensions.

\mysec{\hl{Importance of Considering Various Tile Shapes}}
\hl{The HElayers work} shows that different tile shapes lead to different trade-offs; see a summary of results in Table~II and Table~III of~\cite{helayers}. 
For instance, one tile shape may require more memory but be optimal in terms of execution time, while another shape may trade-off execution time for less memory.
To this end, the framework proposed in~\cite{helayers} uses an optimizer to navigate the space of valid tile shapes and discover the one that is optimized for a given objective function.

\mysec{\hl{Overarching Challenge}}
In the context of \gls{NN} pruning under \gls{HE}, packing the parameters of an \gls{NN} model into tiles raises an important challenge: 
\textit{pruning is effective only if it produces sparsity at the granularity of entire ciphertext tiles},
 and not at that of a single parameter (\eg neuron, or weight) which is the case for plaintext NNs.
This is the exact problem that \FRAME sets out to address.
To this end, we define the term \textit{tile sparsity} as the fraction or percentage of zero tiles (\ie tiles that contain all zero values) out of all the tiles in the NN model.

\subsection{Threat Model}

This paper follows the commonly used threat model for using HE (\eg in \cite{helayers}), but our methods can almost automatically apply to other threat models.
The model involves two parties: a data owner with a pre-trained NN model and  private data samples, and a cloud server for running \gls{HE} inference. 
The data owner generates an HE key-pair, keeps the secret key, and sends the public key and the (possibly encrypted) NN model to the cloud.
Later, the user can securely perform inference by encrypting private samples and uploading them to the cloud, which runs inference and returns the encrypted results for decryption.
During the computation, the cloud learns nothing about the underlying encrypted samples of the user or about the encrypted weights of the model owner, although it does learn the structure of the \gls{NN}, which is provided in the clear by the data-owner.

As the user is also the model-owner, privacy attacks like membership inference over pruned data~\cite{member-attack} are not applicable.
For other threat models, the mitigation technique in~\cite{member-attack} can be applied in our case.
We assume secure protocols for inter-party communication, such as TLS~1.3, and consider computationally-bounded and semi-honest adversaries who faithfully execute protocols.
We modify the data arrangement before encryption without impacting the semantic security of the underlying HE scheme. 
We use 128-bit security in our experiments.


\section{The \FRAME Framework}\label{sec:methodology}

We describe our framework 
and its confidentiality implications in this section.
We provide a description of integrating \FRAME as part of HElayers~\cite{helayers} in App.~\ref{sec:exp-setup}.

\subsection{Prune, Pack, Permute and Expand Methods}\label{sec:methods}
We propose several schemes that prune, re-train and pack an NN model before deploying it.
We list these in Fig.~\ref{fig:ourmethods}. 
Each method starts by training a \gls{NN}; it could train from scratch or start with a pre-trained model. 
We then prune its neurons, weights, or channels, based on some criterion. 
We consider a few standard methods that we summarize in Table~\ref{tab:methods}. 
In the rest of the paper, we name the various pruning techniques using the convention \textit{\{scope/criterion/target\}}.

Our pruning methods accept a \textit{pruning fraction} as input, which is the fraction of the target (weight/neuron/channel) to prune.
We do not consider bias pruning, as biases are generally a small fraction of the parameters in the network.
The pruning criteria we consider include \textit{(i)} uniform-random pruning, \ie randomly pruning neurons/channels or setting weights to zeros, and \textit{(ii)} pruning based on a threshold, \ie the L1-norm of a weight, weights in a channel, or the input/output weights of a neuron.
We  describe  additional criteria for P2T, P4, and P4E later.

The scope of pruning indicates whether the pruning is done layer-by-layer, or all at once. 
When using the random criterion, the scope does not have an effect. 
However, it has a major effect when considering, \eg L1-based pruning, particularly if there is a large variance in the weights across the different layers. 
Here, if we prune some fraction of the network, it might be the case that only the initial layer weights get pruned.
To summarize, we consider five pruning configurations based on valid combinations of the parameters in Table~\ref{tab:methods}, \viz \textit{Lc/L1/Wei}, \textit{Gl/L1/Wei}, \textit{-/Rnd/Wei}, \textit{Lc/L1/Neu} (or \textit{Lc/L1/Chan})\footnote{\textit{Gl/L1/Neu} and \textit{Gl/L1/Chan} were not evaluated since PyTorch limits the scope of global pruning to unstructured methods only.} and \textit{-/Rnd/Wei}.

\begin{figure}[t]
    \centering\scriptsize
    \begin{align*}
        & \operatorname{\meth{P2}}:\hspace{-10pt} & & \operatorname{Train} \rightarrow \operatorname{\meth{P}rune} & & \rightarrow \operatorname{Retrain} \rightarrow \operatorname{\meth{P}ack} \\[0pt]
        & \operatorname{\meth{P2T}}:\hspace{-10pt} & & \operatorname{Train} \rightarrow \operatorname{\meth{P}rune\textsuperscript{\meth{t}ile}} & & \rightarrow \operatorname{Retrain} \rightarrow \operatorname{\meth{P}ack} \\[0pt]
        & \operatorname{\meth{P3}}:\hspace{-10pt} & &  \operatorname{Train} \rightarrow \operatorname{\meth{P}rune} \rightarrow \operatorname{\meth{P}ermute} & & \rightarrow \operatorname{Retrain} \rightarrow \operatorname{\meth{P}ack}\\[0pt]
        & \operatorname{\meth{P3E}}:\hspace{-10pt}  & & \operatorname{Train} \rightarrow \operatorname{\meth{P}rune} \rightarrow \operatorname{\meth{P}ermute} & \rightarrow \operatorname{\meth{E}xpand} &\rightarrow \operatorname{Retrain} \rightarrow \operatorname{\meth{P}ack} \\[0pt]
        & \operatorname{\meth{P4}}:\hspace{-10pt} & &  \operatorname{Train} \rightarrow \operatorname{\meth{P}rune} \rightarrow \operatorname{\meth{P}ermute} \rightarrow \operatorname{\meth{P}rune\textsuperscript{pack}}\hspace{-5pt} & & \rightarrow \operatorname{Retrain} \rightarrow \operatorname{\meth{P}ack} \\[0pt]
        & \operatorname{\meth{P4E}}:\hspace{-10pt}  & & \operatorname{Train} \rightarrow \operatorname{\meth{P}rune} \rightarrow \operatorname{\meth{P}ermute} \rightarrow \operatorname{\meth{P}rune\textsuperscript{pack}}\hspace{-5pt} & \rightarrow \operatorname{\meth{E}xpand} & \rightarrow \operatorname{Retrain} \rightarrow \operatorname{\meth{P}ack} \\[0pt]
    \end{align*}
    \vspace{-32pt}
    \caption{Pruning schemes composed of prune, permute, expand, and pack methods.} 
    \label{fig:ourmethods}
\end{figure}

\begin{table}[t]
\centering
\caption{Scope, criterion and target of pruning in each of the pruning schemes. See explanation in Section~\ref{sec:methods}.}
\label{tab:methods} \vspace{-5pt}
\resizebox{.8\columnwidth}{!}{
    \begin{tabular}{|l||l|l|}\hline\footnotesize
    \multirow{3}{*}{P2T}         & Scope     & Local (\textsf{Lc}), Global (\textsf{Gl})                                                                                                                    \\ \cline{2-3} 
                                 & Criterion & \begin{tabular}[c]{@{}l@{}}Average/Maximum/Minimum of tile \\ (T-Avg/T-Max/T-Min)\end{tabular}                                             \\ \cline{2-3} 
                                 & Target    & Weight (\textsf{-}) [tile granularity]                                                                                                                                          \\ \hline\hline
    \multirow{3}{*}{P2, P3, P3E} & Scope     & Local (\textsf{Lc}), Global (\textsf{Gl})                                                                                                                    \\ \cline{2-3} 
                                 & Criterion & L1 (\textsf{L1}), Rand (\textsf{Rnd})                                                                                                                        \\ \cline{2-3} 
                                 & Target    & Weight (\textsf{Wei}), Neuron (\textsf{Neu}) / Channel (\textsf{Chan})                                                                                                                 \\ \hline\hline
    \multirow{3}{*}{P4, P4E}     & Scope     & Local (\textsf{Lc}), Global (\textsf{Gl}) {[}1st and 2nd prune{]}                                                                                            \\ \cline{2-3} 
                                 & Criterion & \begin{tabular}[c]{@{}l@{}}L1 (\textsf{L1}), Rand (\textsf{Rnd}) {[}1st prune{]}, threshold fraction of zeros\\ in a tile, above which the tile is pruned {[}2nd prune{]}\end{tabular} \\ \cline{2-3} 
                                 & Target    & Weight (\textsf{Wei}), Neuron (\textsf{Neu}) / Channel (\textsf{Chan}) {[}1st prune{]}                                                                                                 \\ \hline
\end{tabular} 
}\vspace{-10pt}
\end{table}

\begin{figure*}[t]
    \centering
    \includegraphics[width=.9\linewidth]{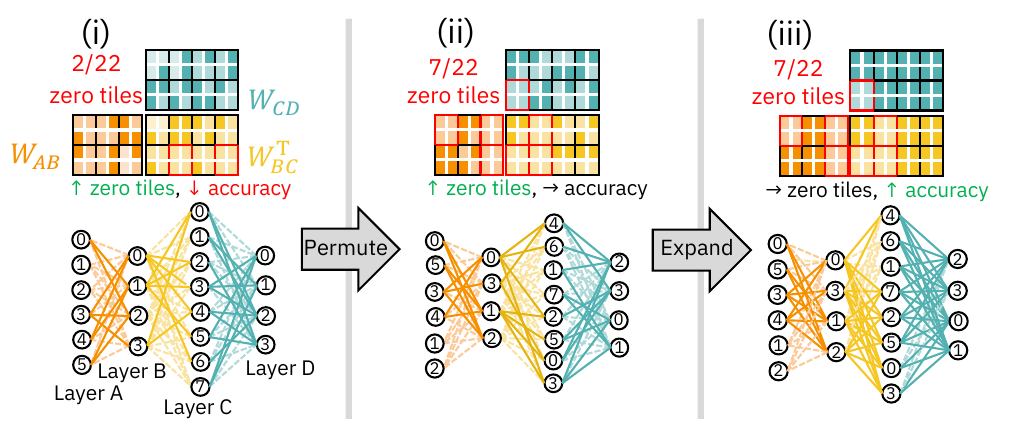}
    \vspace{-10pt}
    \caption{Illustration of permutation and expansion for the P3E scheme when considering a $4$-layer network with $6$,$4$,$8$,$4$ neurons in layers A-D, resp.
    We divide the weight matrices into $2 \times 2$ tiles and prune $54/88=61\%$ of the weights. 
    The pruned \gls{NN} \textit{(i)} has a tile sparsity of only $2/22 \approx 10\%$ zero tiles. 
    Permutation \textit{(ii)}, improves it to $7/22=35\%$. 
    Expansion \textit{(iii)} (with re-training) restores most of the accuracy loss.}
    \label{fig:weight-prune}\vspace{-10pt}
\end{figure*}


All of the strategies, except for P2T, first perform pruning using one of the pruning configurations listed above. 
P2T uses a tile-based pruning configuration that is distinct from these. 
Here, we first pick a tile shape and split every matrix into tiles, as illustrated in 
Fig.~\ref{fig:tiletensors}.
For every tile, we compute the minimum/ maximum/average metric of its absolute values and prune a fraction of the tiles with the lowest values of this metric.
We refer to these options under the (reduction) criteria in Table~\ref{tab:methods} and denote this pruning as \underline{\bf P}rune\textsuperscript{\underline{\bf t}ile} in Fig.~\ref{fig:ourmethods}.
This forms our adaptation of the Hunter~\cite{hunter} scheme in the context of tile tensors~\cite{helayers}. 

P2T prunes complete tiles (\ie packing-aligned pruning) right away and, therefore, there is no need to perform further steps, such as \underline{\bf P}ermute, \underline{\bf P}rune\textsuperscript{pack}, or \underline{\bf E}xpand.
However, a major disadvantage of P2T is that because each tile may harbor a wide range of weights, important weights may get pruned out.
Intuitively, a non-packing-aligned pruning scheme may be more efficient in terms of removing the ``unimportant'' weights. 
However, the pruned values are not necessarily organized in a way that is packing-friendly and would lead to cancellation of tile operations. 
Therefore, we propose additional steps to mitigate this.

The permute and expand operations are illustrated in Fig.~\ref{fig:weight-prune}.
The \textit{permute} operation permutes the rows and columns of the weight matrices after the pruning operation to essentially congregate the zero elements together. 
The detailed algorithm that we deploy is described in Section~\ref{sec:permute}. 
The \textit{expand} operation is a partial reversal of the pruning operation, where we search for tiles that contain both zeros and non-zeros, and we un-prune the zero elements inside them.
The motivation behind this action lies in the motivation for pruning, which is to reduce the number of active tiles;
if a tile is not reduced, \ie it has non-zero elements, then we cannot ignore it, and because we do not gain any performance benefits, it is best to fully utilize its elements to improve the model accuracy.

The aforementioned strategies  form the foundations of the schemes \hl{that are abbreviated as} P2, P2T, P3, and P3E in Fig.~\ref{fig:ourmethods}.
To complete the set, we construct two more schemes, called P4 and P4E. In P4, instead of expanding the model as in P3E, we perform a second packing-aware pruning step to remove tiles that contain ``mostly'' zeros. 
These tiles are selected based on whether they contain more than a threshold fraction of zeros (an additional criterion for this scheme in Table~\ref{tab:methods}).
P4E adds an expand operation after the second pruning step in P4.

\begin{figure}[t]
    \centering
    \includegraphics[width=1\linewidth]{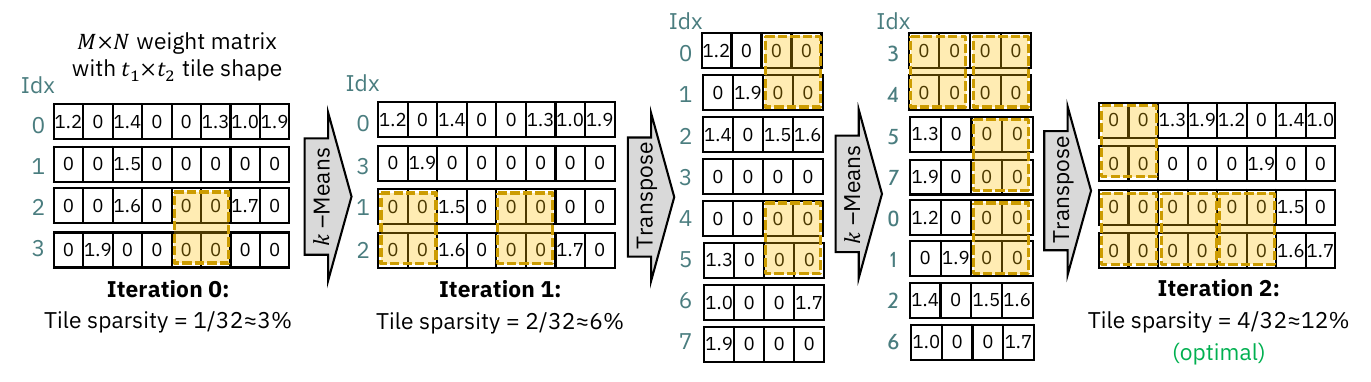}\vspace{-10pt}
    \caption{Permutation of a single $4$$\times$$8$ weight matrix considering \hl{tiles of shape $2$$\times$$2$}, with zero tiles highlighted.
    We show the algorithm on the weight matrix instead of the mask matrix, for illustration.
    Here, two \textit{k}-means iterations are sufficient to reach the solution with the maximum tile sparsity (optimality tested using exhaustive search).}
    \label{fig:single-layer}\vspace{-15pt}
\end{figure}

\subsection{Layer Co-Permutation for Improved Pruning}\label{sec:permute}

We propose an efficient way of increasing the tile sparsity of an \gls{NN} model, given a pruned NN and a tile shape.
We first discuss the algorithm for a single weight matrix (\gls{FC} layer) and then extend it to \glspl{NN} with multiple FC and Conv layers.

Given a pruned matrix, one may ideally desire shuffling the values to arrange the zeros and non-zeros into separate tiles.
However, it is not possible to arbitrarily swap two values without changing the function that the matrix encodes.
\textbf{Our key insight} is that one can, instead, shuffle the rows and columns of the matrix to improve the tile sparsity, \textit{without affecting the functionality.}
For an FC layer, it is easy to understand row-column permutations as reordering the order of the neurons of the layers that neighbor the weight matrix, \eg in Fig.~\ref{fig:weight-prune} \textit{(i)}--\textit{(ii)}.
One na\"ive way to obtain a permuted matrix that maximizes the tile sparsity with this method is through exhaustive search. 
However, this  has a complexity of $\mathcal{O}(M!N!)$ for an $M$$\times$$N$  matrix, which is prohibitive for large matrices.

To make this problem tractable, we propose to permute the rows and columns based on iterative clustering algorithms. 
We propose a variant of $k$-means, which we illustrate in Fig.~\ref{fig:single-layer}.
Specifically, we split the weight array into rows, convert them to binary mask vectors (non-zeros become 1s) and perform \textit{k}-means ($k$$=$$\lceil\frac{t_1}{M}\rceil$) using a distance function that, for two mask vectors (or, points) \texttt{a\textsubscript{i}} and \texttt{b\textsubscript{i}}, computes the number of non-zero groups of size $\lceil\frac{t_2}{N}\rceil$ in the vector \texttt{(a\textsubscript{i}\&b\textsubscript{i})}.
We term this as \textit{Grouped Hamming} distance; App.~\ref{app:p4e_meta_pr_frac} shows a comparison with a few other variants.
Further, we implement a balancing scheme that reassigns points from centroids with $>$$t_1$ points to centroids that have the minimum distance to the point.
After running $k$-means, the reordered points are transposed, $\{t_1, M\}$ and $\{t_2, N\}$ are swapped, and the process is repeated till convergence.

\begin{figure}[t]
    \centering
    \includegraphics[width=\linewidth]{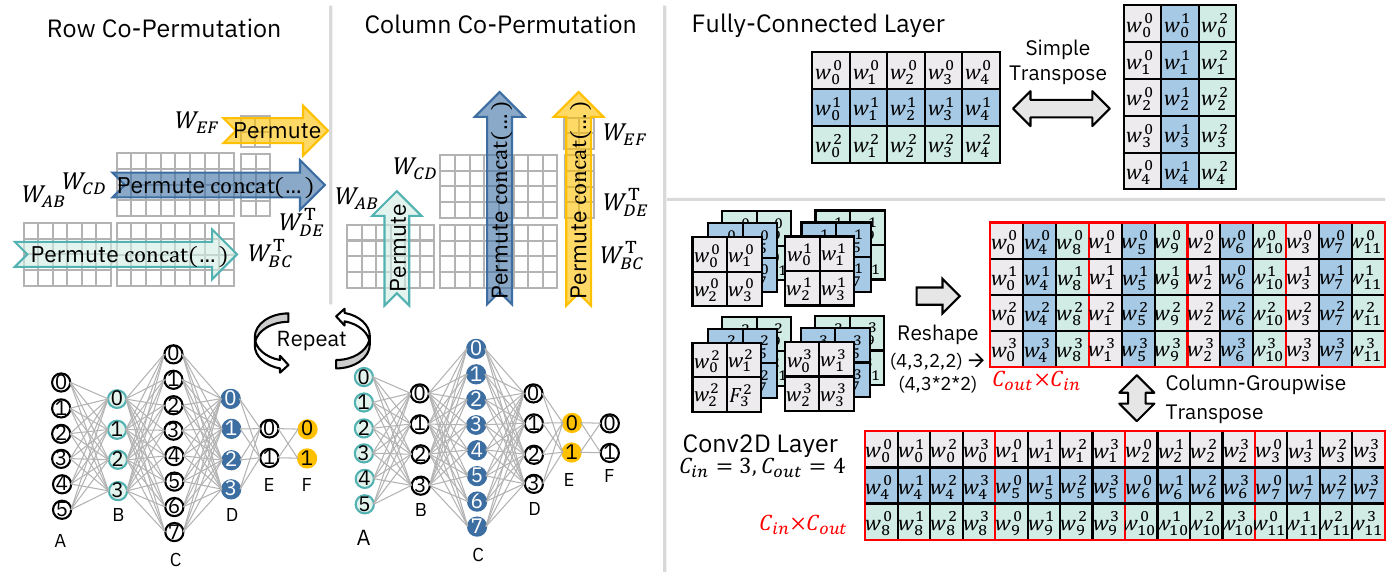}\vspace{-10pt}
    \caption{\textbf{Left.} Illustration of weight co-permutation in a multi-layered FC-only network with weight matrices $W_{AB}, \ldots, W_{EF}$. The row and column permutation phases separately permute three different sets of matrices, where the permute operations correspond to permuting the highlighted neurons. \textbf{Right.} Illustration of weight transposition in the context of FC \textit{(top)} and Conv \textit{(bottom)} layers.}
    \label{fig:multi-layer}\vspace{-15pt}
\end{figure}
\mysec{Multi-Layer Permutation} 
The issue with permuting each layer of the network \textit{independently} using the aforementioned technique is that a new ``permute'' layer must be added after each of the original layers for functional correctness of the network, which can easily remove any benefits from improving the tile sparsity.
We obviate the need for these ``permute'' layers by modifying our technique to permute connected layers in tandem.
We refer to this as layer \textit{co-permutation}.

Consider the network in Fig.~\ref{fig:multi-layer}. 
Here, shuffling the neurons in layer B implies co-permuting the rows of $W_{BC}^T$ and the rows of the preceding weight matrix, $W_{AB}$, as opposed to permuting each matrix separately. 
Similarly, permuting the neurons of layer D translates to permuting $concat(W_{CD},W_{DE}^T)$.
In the next iteration, we permute the columns of the re-grouped sets of weights, which corresponds to permuting the neurons in layers A, C, and E.
This process is repeated till convergence.
As is evident, this technique essentially discovers a re-arranged network such that the packing method can discard the maximum number of ciphertexts (of tiles) that contain only zeros, thus benefitting both memory and latency, and without affecting the model (and, thus, accuracy/loss).

\mysec{Extension to Conv Layers} In our actual implementation, we transpose each weight matrix before applying row-wise \textit{k}-means to groups of matrices, rather than switching between row- and column-wise \textit{k}-means each iteration.
While this transpose operations implies a simple transposition for FC layers (Fig.~\ref{fig:multi-layer} top-right), the technique is more involved when applied to 2D convolution (Conv) layers.
Precisely, the Conv layer is reshaped to reduce it to a 2D tensor and each group of $C_{out}$$\times$$C_{in}$ (or,  $C_{in}$$\times$$C_{out}$, for alternate iterations) sub-matrix is transposed within the larger tensor.
At a higher level, this permutes the channels of the activations consumed and produced by the Conv layer.
Note that the reshape and transpose operations are done prior to model deployment (during training) and, thus, do not add any overhead during inference.

\mysec{Permuting Activations}
{Along with the permuted model, our algorithm produces two permutation matrices, $P_{in}$ and $P_{out}$, which are multiplied with the input and output, resp., at the client end, in the clear.
The overhead of these two plaintext operations is negligible compared to \gls{HE} computation.
For \glspl{CNN}, the presence of pooling after the last Conv layer renders the co-permutation of a Conv and FC layer-pair difficult.
Therefore, we append a single ``permute'' ({MatMul}) layer before the first FC layer, which is not expected to have much overhead as it is extremely sparse (\eg $\sim$$97\%$ zero 16$\times$16 tiles for AlexNet). Alternatively, one can fuse it with the first FC layer.}

\subsection{Privacy Considerations}\label{sec:sec}
The confidentiality of the model or the user data is agnostic to the use of \FRAME, because the pruning phase happens in plaintext before uploading any data to the server. 
Once the data is encrypted, its security relies on the semantic security of the underlying \gls{HE} algorithm, 
which means that the server can perform any operation on the ciphertexts without revealing the underlying plaintexts. 
Another advantage of using a non-interactive design is that it relies only on the \gls{HE} primitive and does not rely on additional cryptographic primitives such as \glspl{GC}, \gls{OT}, or \gls{SS}. 

By pruning a model, we only reduce the amount of memory, and computation that a server needs to perform; we do not introduce new cryptographic mechanisms. 
Following~\cite{hunter}, and most prior works in \gls{HE}, we assume that no meaningful leakage happens from using a pruned model, except for the model architecture, which includes the position of  zero tiles and is visible with \gls{HE}.
With that being said, we believe that further research in this area will be valuable.


\section{Experimental Setup}\label{sec:setup}
\begin{figure}[t]
    \centering
    \includegraphics[width=.9\linewidth]{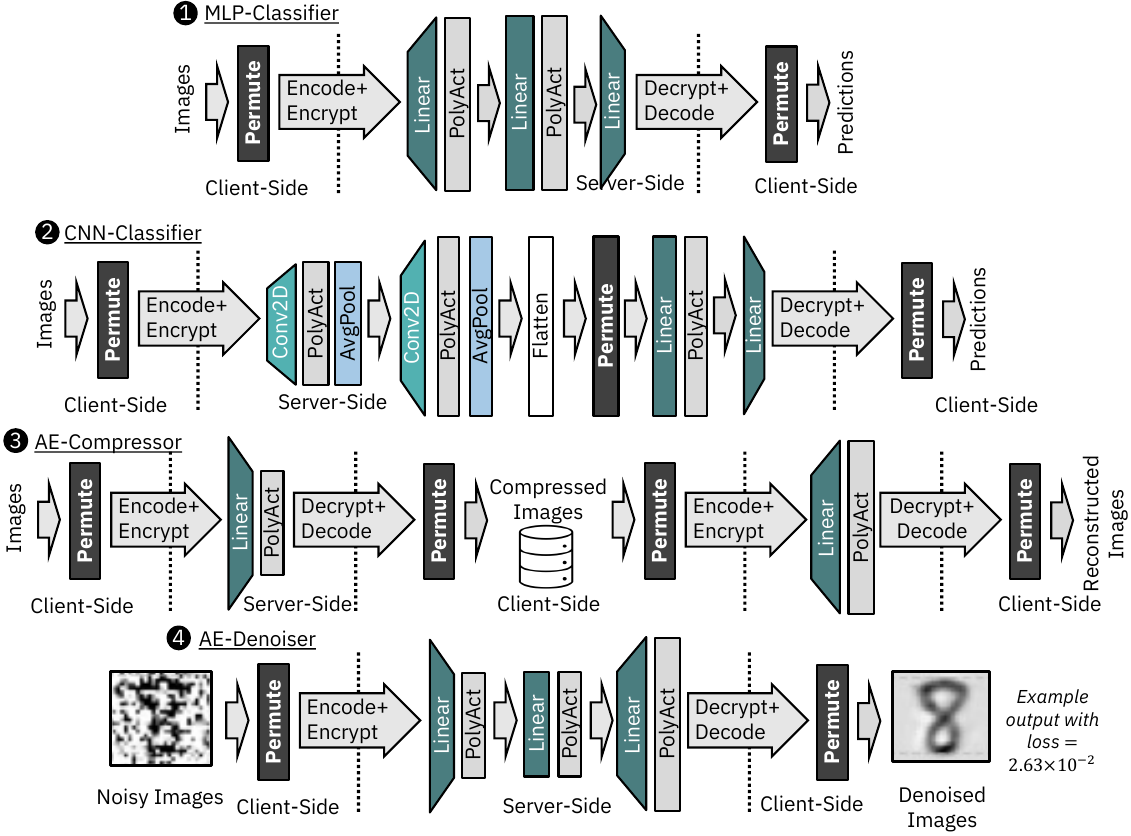}\vspace{-10pt}
    \caption{Privacy-preserving neural network architectures evaluated in this work. To help with an intuitive understanding of the test loss, for the AE-Denoiser, we included an example input and output image that was produced by a model with 70\% tile sparsity.}
    \label{fig:apps}\vspace{-15pt}
\end{figure}

\mysec{Network Architectures} We evaluate our methods on the four applications in Fig.~\ref{fig:apps} and list their parameters in App.~\ref{sec:exp-setup}.
\Glspl{MLP} and \glspl{CNN} are commonly used for classification tasks using \gls{FC} (linear) layers.
CNNs are specifically designed to process pixel data and additionally use 2D convolution layers.
Autoencoder networks can learn to compress and reconstruct data, through dimensionality reduction.
They are also used for image denoising by learning noise from images and then eliminating it for new images.
Following~\cite{baruch2022methodology}, we substitute \textit{(i)} ReLU activations with polynomial activations, and \textit{(ii)} max-pooling with average-pooling, to make the \glspl{NN} amenable to HE math.

As an additional demonstrator, we showcase our P4E technique on a 21-layer HE-friendly AlexNet, as in \cite{helayers}, trained using trainable activations as in~\cite{baruch2022methodology}.

\mysec{Datasets}
We experiment using the following datasets:
MNIST that has 60,000 $28$$\times$$28$ images of handwritten digits, 
CIFAR-10, a collection of 50,000 small color images of everyday objects, and
SVHN, an image dataset with $>$600,000 house number images from Street View~\cite{netzer2011reading}.
Following \cite{baruch2021fighting,helayers}, the AlexNet network is trained on the COVIDx CT-2A dataset~\cite{gunraj2020covidnet} containing 194,922 chest CT slices. 

\mysec{Tile Shapes}\label{sec:tile_shapes}
Without loss of generality, we perform our experiments with tile tensors of dimensions $[t_1, t_2(=t_1), t_3]$, where $t_1,t_2$ correspond to the width and height of tiles in the weight matrices, and $t_3$ is the batch size.
We fix the ciphertext poly-degree ($N$) to $32{,}768$, which allows it to hold $\frac{N}{2}=16{,}384$ values.
We set $t_1 \in \{8,16,32,64\}$ and adapt $t_3$ so that ($t_1\cdot t_2\cdot t_3=\frac{N}{2}$).
The choice of tile shape trades-off latency, throughput, and memory, as described in~\cite{helayers}.

\mysec{\hl{Evaluation Metrics}}
We vary the pruning fraction in 0.5--5\% increments
and plot the tile sparsity, \ie \% of resulting zero ciphertexts after packing the \gls{NN} weights, against the cross-entropy test loss (or accuracy) of the  \gls{NN}. 
We also report the inference latency and memory requirement using HElayers. 
We define an upper limit for the inference test loss/accuracy degradation over the unpruned \gls{NN} as 2.5\% .
However, this is left as a parameter for the model-owner to decide. 

\mysec{\hl{Platforms}}
App.~\ref{sec:exp-setup} lists the platforms used for our experiments.
\section{Evaluation}\label{sec:exp}
We present the evaluation of our best pruning methods for each of the networks in Fig.~\ref{fig:apps} and compare them to our baseline~\cite{hunter}. 
We then report results from deploying our best method on a \gls{PPML} inference scenario using HElayers~\cite{helayers}.


\subsection{Analysis of Our Pruning Schemes}\label{sec:pruning_scheme_analysis}
\begin{figure}[t]
    \centering
    \includegraphics[width=\linewidth]{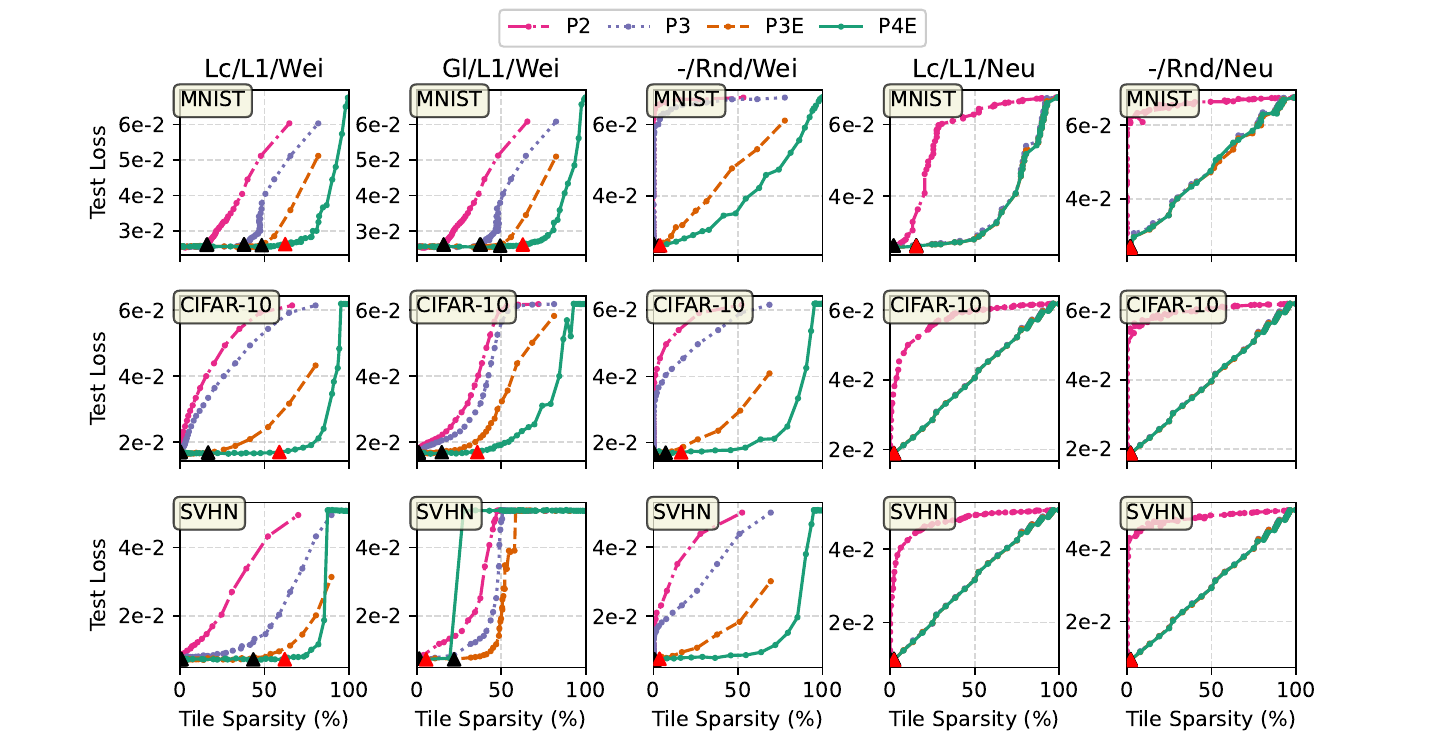}\vspace{-10pt}
    \caption{Evaluation of our  techniques  on the autoencoder-based denoiser \gls{NN} for $16$$\times$$16$ tiles. Columns show different pruning \textit{scopes/criteria/targets} while rows show different datasets.
    \ulg (P4E)/$\blacktriangle$ (others) show the \gls{NN} that, for each technique, has the highest tile sparsity (\% of zero tiles) with $\leq2.5\%$ worse inference loss over the unpruned \gls{NN}.
    }
    \label{fig:method_comparison}\vspace{-10pt}
\end{figure}

We now compare our P2, P3, P3E, and P4E pruning schemes for each of our \textit{\{scope/criterion/target\}} configurations (Table~\ref{tab:methods}).
Fig.~\ref{fig:method_comparison} compares these for four tile shapes, on the basis of test loss/accuracy and tile sparsity, for the AE-Denoiser \gls{NN}.
We note that an ideal pruning scheme should maximize the tile sparsity of the \gls{NN} model for minimum degradation in test loss/accuracy. 

We present our analysis in the order of the worst to the best pruning scheme.
First, we note that P2, which is standard for plaintext NN pruning, fails for network pruning under HE across all networks, datasets, and schemes, thus motivating the need to consider improving the tile sparsity instead of regular sparsity (Section~\ref{sec:intro}).
In particular, neuron pruning (\textit{\{Lc/L1/Neu\}} and \textit{\{-/Rnd/Neu\}}) with P2 quickly break down in terms of loss, without generating more than a few zero tiles.
After permutation (P3/P3E/P4E), the network becomes almost fully-dense, \ie except for the tiles at the right or bottom edge of the weight tensor, which are taken care of by expansion (P3E/P4E).
With P3, our methods improve the number of zero tiles, but not the test loss.

The \textit{-/Rnd/Wei} configuration, which randomly prunes weights in the network, is oblivious to the importance (value) of the weights and, thus, is not as robust as L1-norm--based pruning.
Permutation (P3/P3E/P4E) helps, but not by much, as the distribution after permutation is also uniform-random.
P4E, with the combination of two pruning rounds with permutation and expansion, performs the best, but as we will see, other  schemes outperform random pruning.

The local and global L1-norm based weight pruning schemes, \textit{Lc/L1/Wei} and \textit{Gl/L1/Wei}, demonstrate the best overall trade-off of test loss and tile sparsity, as shown with the colored markers.
We also observe that permutation is more effective for the local-scope case than for the global scope.
For this \gls{NN}, local pruning outperforms global pruning, however this is not always the case, as we will see.
Overall, P4E is our most effective scheme, as it is able to push the  pruning envelope beyond just prune+permute with an additional layer of \textit{careful} pruning.
As a trade-off, P4E introduces a new hyperparameter, \ie the threshold for Prune\textsuperscript{pack}, which we set to 93.8\% based on a set of experiments (App.~\ref{app:p4e_meta_pr_frac}).


\begin{figure}[t]
    \centering
    \includegraphics[width=\linewidth]{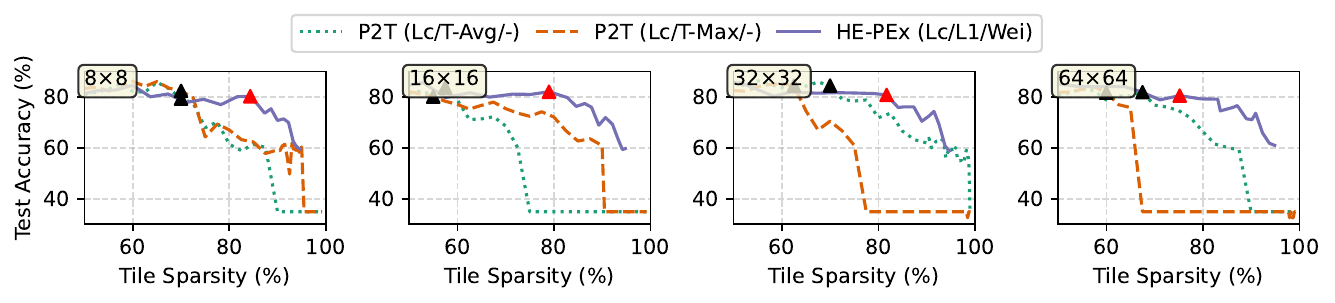}\vspace{-10pt}
    \caption{Inference accuracy \vs tile sparsity (\% of zero tiles) comparison using our best pruning technique (P3E for Conv and P4E for \gls{FC} with Lc/L1/Wei, labeled \FRAME) against P2T variants for a 21-layer AlexNet \gls{NN} on the COVIDx CT-2A dataset. The tile shapes are shown in the top-right corner of each subplot. \ulg (\FRAME)/$\blacktriangle$ (P2T) shows the network configuration that has the highest tile sparsity (\% of zero tiles) within at most 2.5\% worse accuracy compared to the original (unpruned) network.}
    \label{fig:pruning_comparison}\vspace{-15pt}
\end{figure}

\mysec{Best Overall Scheme}
Our best scheme, referred to as HE-PEx henceforth, uses P4E for FC layers, but P3E for Conv layers; this is because Conv layers are more sensitive to the second pruning round (Prune\textsuperscript{pack}). 
Alternatively, hyperparameter search techniques can be used to explore different values of the pruning threshold for Prune\textsuperscript{pack} to mitigate this, but this is not explored in our paper.

\mysec{Local \vs Global Pruning}
Prior work~\cite{blalock2020state} shows that global pruning outperforms local pruning by avoiding new hyperparameters associated with {per-layer} pruning fractions. 
However, recent works, such as~\cite{wang2020picking}, have demonstrated that global pruning may lead to \textit{layer-collapse} that breaks the network, as we can clearly see in the P4E plot for SVHN with \textit{Gl/L1/Wei} configuration in Fig.~\ref{fig:method_comparison}.
We observe the efficiency of local \vs global pruning for tile sparsity to be dependent on the architecture and the dataset in our experiments.

\begin{figure}[t]
    \centering
    \includegraphics[width=\linewidth]{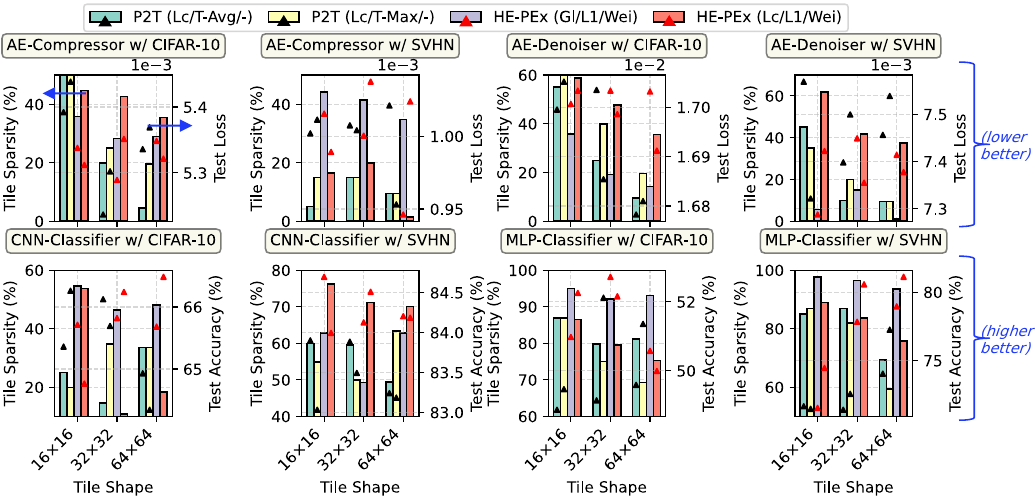}\vspace{-10pt}
    \caption{\textit{Left axis.} Best tile sparsity (\% of zero tiles) for within-2.5\% degradation in inference accuracy/loss over the unpruned \gls{NN}. \textit{Right axis.} Absolute inference accuracy/loss \vs tile shape comparison using our best pruning technique (P3E for Conv and P4E for \gls{FC}, labeled \FRAME) against P2T variants. Bars show zero tile \% and  \ulg (\FRAME)/$\blacktriangle$ (P2T) markers show the corresponding inference accuracy/loss. Global-scope P2T variants are not shown as they perform much worse than local-scope variants.}
    \label{fig:all_apps_analysis}\vspace{-15pt}
\end{figure}

\subsection{Comparison with P2T Strategies}\label{sec:p2t_comp}

We now compare our best pruning method with our adaption of Hunter~\cite{hunter} to a non-interactive solution, which we refer to as P2T. 
We construct four variants of P2T: \textit{Gl/T-Avg/-}, \textit{Gl/T-Max/-}, \textit{Lc/T-Avg/-} and \textit{Lc/T-Max/-}.
All of these use packing-aware pruning as the first step, with the scope and reduction criterion specified by the first two parameters in the scheme name (Section~\ref{sec:methods}), resp.

We first report the results for the HE-friendly AlexNet network trained on the COVIDx CT-2A dataset, in Fig.~\ref{fig:pruning_comparison}.
Here, the local-scope variants outperform the global-scope schemes, as is true of P4E from Section~\ref{sec:pruning_scheme_analysis}, and thus we do not plot those.
Our best scheme achieves a tile sparsity of \textbf{75--84\%} with, at worse, a \textbf{2.5\%} degradation in inference accuracy over the unpruned \gls{NN}. 
The efficacy of the P2T schemes, however, degrades as the tile size increases. 
This is because the variance of magnitudes of weights in each tile grows with the size of the tile, \ie the larger the tile, the greater the probability that it holds a set of ``important'' weights along with ``unimportant'' ones.
Pruning all of these weights at once in P2T leads to severe accuracy degradation, since important weights are pruned out. 
These results showcase the benefits of our prune, permute and expand methods, in lieu of blind packing-aware pruning schemes, such as P2T.

We now report a summary of our results across the four \glspl{NN} in Section~\ref{sec:exp}.
Fig.~\ref{fig:all_apps_analysis} shows the best tile sparsity achieved for $<$$2.5$\% worse test accuracy (or loss) on CIFAR-10 and SVHN.
Overall, we observe that HE-PEx improves the tile sparsity of each \gls{NN} by \textbf{48\%}, \textbf{57\%}, and \textbf{104\%} over P2T across the three tile sizes, resp. 
In the few cases where there is a smaller improvement, our methods still produce models that are more accurate than those obtained using P2T.

\mysec{Key Takeaway}
Our methods produce NNs with tile sparsities of \textbf{95}\% for the MLP-Classifier, \textbf{61}\% for the CNN-Classifier, \textbf{41}\% for the AE-Compressor, and \textbf{47}\% for the AE-Denoiser, averaged over the two datasets, for $<$2.5\% degradation in inference accuracy/loss over the unpruned network.
These \glspl{NN} have drastically smaller latency and memory requirements, which we will see in Section~\ref{sec:lat_mem_analysis}.

\subsection{Impact of Permutation}\label{sec:perm_analysis}
\begin{wrapfigure}{r}{0.6\textwidth}\vspace{-10pt}
  \begin{center}
    \includegraphics[width=0.58\textwidth]{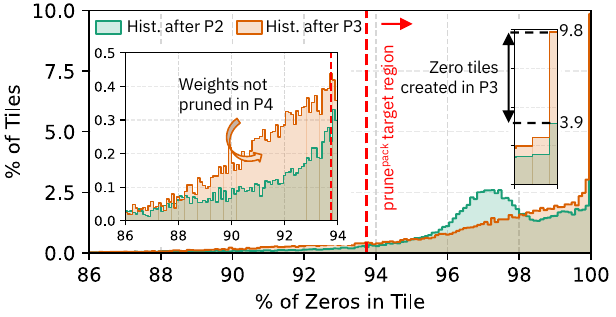}
  \end{center}
  \vspace{-25pt}
  \caption{Histograms of zero values in $32$$\times$$32$ tiles of the layers of AlexNet trained on the COVIDx CT-2A dataset, after performing Lc/L1/Wei pruning.}
  \label{fig:zerohists}
\end{wrapfigure}

To visualize the benefits of permutation, we plot histograms of the percentage of tiles that have a certain percentage of zeros in them.
This is shown for AlexNet with \textit{Lc/L1/Wei} pruning of 97.5\% of the weights, in Fig.~\ref{fig:zerohists}. 

\setlength{\intextsep}{1pt}

As shown in green in the figure, after pruning these weights as part of the first prune step, we see a Gaussian distribution centered at 97.5\%, with a squashed tail.
Here, only 3.9\% of the tiles have \textit{all} zeros. 
After  permutation, (orange histogram), we observe a two-fold set of benefits.
First, and more obviously, the weight tensors are rearranged so that more zeros are clustered together, thus increasing the tile sparsity from 3.9\% to 9.8\%.
Second, trained non-zero weights have migrated from  sparse tiles into denser tiles.
Particularly, tiles with density less than the Prune\textsuperscript{pack}  threshold in P4 have increased (see the 86--94\% region in Fig.~\ref{fig:zerohists}), which  leads to better inference accuracy after re-training.
As one would expect, these benefits become more prominent with larger $t_1,t_2$ dimensions of the tile.

\vspace{-2pt}
\subsection{Impact on Latency and Memory}\label{sec:lat_mem_analysis}
\begin{figure}[t]
    \centering
    \includegraphics[width=.9\linewidth]{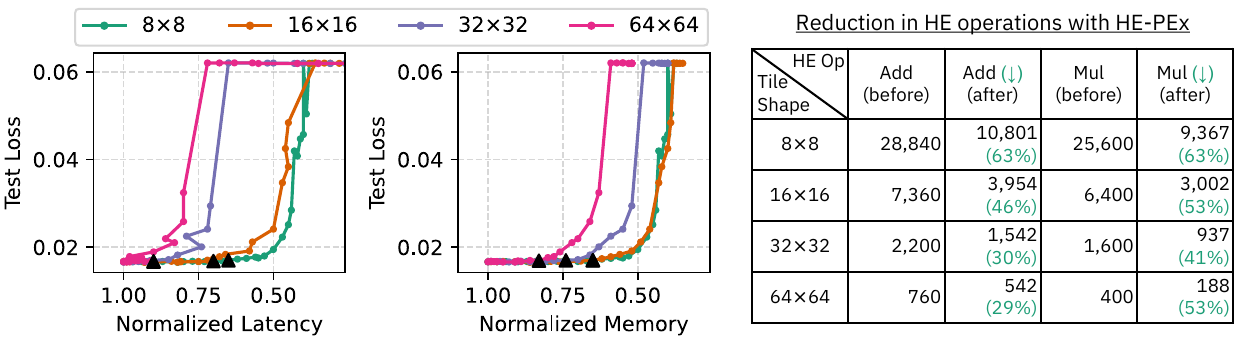}\vspace{-10pt}
    \caption{Trade-offs of latency and memory requirement \vs test loss (\textit{left}/\textit{center}) and {HE operation count before and after pruning} (\textit{right}) for the AE-Denoiser \gls{NN} with the CIFAR-10 dataset for different tile shapes, measured using HElayers with the SEAL HE library. $\blacktriangle$ markers show the \gls{NN} configurations that have the smallest latency (left) or memory (center) for a loss degradation of at worst 2.5\% over the unpruned \gls{NN}.}\vspace{-15pt}
    \label{fig:helayerStats}
\end{figure}

Fig.~\ref{fig:helayerStats} (left) shows the normalized latency and memory reduction, along with the test loss of the network, when running inference HE using HElayers integrated with \FRAME, for the AE-Denoiser on CIFAR-10.
Here, we vary the tile shape as $t_1$=$t_2$=$\{8,16,32,64\}$ and fix the batch dimension, $t_3$, to $\{256,64,16,4\}$, resp., (Section~\ref{sec:tile_shapes}). 
Our results show memory and latency reductions of \textbf{10--35\%} and \textbf{17--35\%}, resp., compared to the unpruned model, within a \textbf{2.5\%} degradation in test loss, which increase to become up to \textbf{41\%} if a \textbf{5\%} degradation is acceptable.
The benefits are higher for smaller tile shapes, as the tile sparsity is greater for smaller tiles.
However, smaller tile shapes require a large $t_3$, and therefore a large batch size of the inputs~\cite{helayers}.
It is possible to zero-pad along the third dimension if there are insufficient inputs per batch; however, it would lead to wasted storage, and therefore, a larger tile shape is preferred in this case.
The final choice of tile shape depends on the  batch size, latency, memory, and/or throughput (see~\cite{helayers}).

Fig.~\ref{fig:helayerStats} (right) shows a breakdown of the number of \gls{HE} additions and multiplications before and after pruning. As expected, pruning reduces \textbf{29--66\%} of the \gls{HE} additions and \textbf{53--67\%} of the \gls{HE} multiplications. 
Our experiments did not show a significant reduction in the number of HE rotations. 
One explanation is that when using tile tensors to perform matrix multiplication, even with pruning, we still need to execute the rotate-and-sum algorithm~\cite{helayers}. 
Another observation is with the right-most point in latency graph for the $8$$\times$$8$ tile shape in Fig.~\ref{fig:helayerStats}; this point corresponds to a 99\% tile sparsity and yet has a normalized latency of 0.37.
The residual latency exists because although we reduce 67\% of the multiplications that are part of the matrix multiplication operations of the \gls{NN}, the \textit{output} of an \gls{FC} layer is unlikely to include zero tiles.
Therefore, we still have to perform the activations, which involve additional multiplications.
\vspace{-5pt}
\section{Conclusion}\label{sec:conc}
This work motivated the challenges associated with employing plaintext pruning techniques for PPML networks and presented a set of methods for efficiently pruning \glspl{NN} for PPML inference under HE.
Our solution, called \FRAME, combines four critical primitives: \textit{pruning}, \textit{permutation}, \textit{expansion}, and \textit{packing}. 
Specifically, we introduced a novel permutation algorithm that rearranges the \gls{NN} weights to improve pruning efficiency, without affecting the accuracy.
We described how \FRAME operates in non-interactive HE inference use cases and integrated it with 
HElayers~\cite{helayers}.
The integrated framework takes as input the unpruned \gls{NN} and produces a deployment-ready pruned \gls{NN} that meets an objective function composed of 
accuracy, latency, and memory requirement.

We demonstrated our  techniques  on a set of four PPML applications with four datasets. 
We adapted a \gls{SotA} pruning technique called Hunter~\cite{hunter} for non-interactive HE and compared our best scheme against theirs, in terms of inference accuracy (or loss) \vs tile sparsity in the pruned \gls{NN}.
Across different tile shapes, \glspl{NN}, and datasets, our framework produced pruned models with \textbf{70\%} larger tile sparsity on average, over Hunter, within an inference accuracy/loss degradation threshold of \textbf{2.5\%} over the unpruned \glspl{NN}.
Our implementation on top of HElayers showed up to \textbf{35\%} reduction in inference latency and memory requirement, over the unpruned \gls{NN}, for a PPML image denoising application.

\appendix
\section{Appendix}

\begin{figure}[t]
    \centering
    \includegraphics[width=\linewidth]{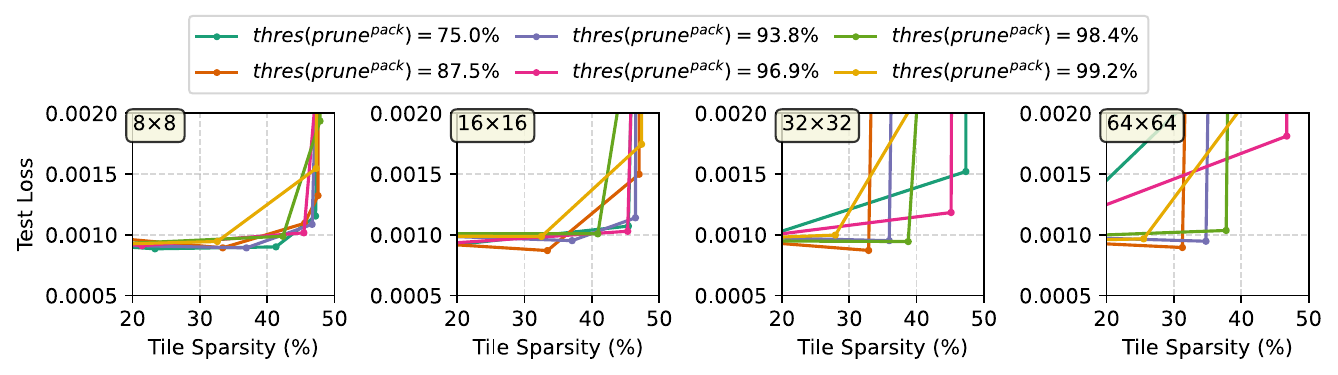}
    \caption{Test loss \vs tile sparsity (\% of zero tiles) for different prune\textsuperscript{pack} thresholds of our P4E scheme, considering four tile shapes using the AE-Compressor with CIFAR-10.
    }
    \label{fig:P4E_sweep}
\end{figure}

\subsection{Additional Evaluation}\label{app:p4e_meta_pr_frac}
\mysec{Selection of P4E prune\textsuperscript{pack} Threshold}
We justify our choice of selecting a threshold for the prune\textsuperscript{pack} step (threshold \% of zeros in the tile, above which the tile is pruned) of P4E by running a sweep and reporting the results in Fig.~\ref{fig:P4E_sweep}, for our AE-Compressor with CIFAR-10.
The behavior is similar for other networks and datasets.
We observe that large values of the threshold do not introduce sufficient new zero tiles to make any meaningful improvement upon P3E.
At the other extreme, too small values of this threshold aggressively remove weights, leading to degradation in  test loss.
We select 93.8\% as our choice here, but posit that our results can be further improved by tuning this as a hyperparameter.

\begin{wrapfigure}{r}{0.37\textwidth}
  \begin{center}\vspace{-10pt}
    \includegraphics[width=0.365\textwidth]{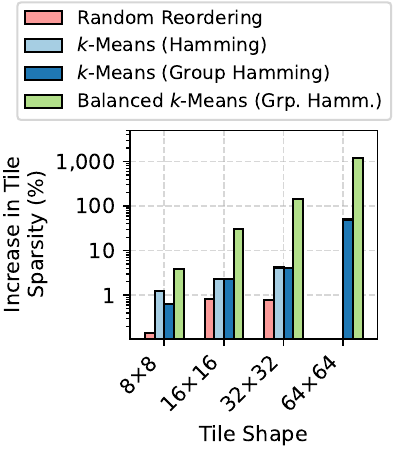}
  \end{center}\vspace{-20pt}
  \caption{Increase in tile sparsity (log scale) with different permutation algorithms for AlexNet on the COVIDx CT-2A dataset, after \textit{Lc/L1/Wei} pruning.}
  \label{fig:app:perm_comp}
\end{wrapfigure}

\mysec{Comparison of Permutation Algorithms}
We compare different permutation algorithms for the permute step in \FRAME.
Fig.~\ref{fig:app:perm_comp} shows the results of AlexNet with \textit{Lc/L1/Wei} pruning of 97.5\% of the weights, which is the same scenario discussed in Section~\ref{sec:perm_analysis}.

\setlength{\intextsep}{0pt}
As expected, ``Random Reordering'' for the row-column co-permutation provides little-to-no benefit, due to the sheer number of possible permutations of the entire network.
Interestingly, $k$-Means alone only improves the benefits to $<$10\%.
Moreover, the use of a ``Grouped Hamming'' variant (Section~\ref{sec:permute}) does not show significance for the smaller tile shapes.
However, as seen in the $64$$\times$$64$ case, grouped Hamming distance is critical for grouping zeros together for large tile shapes.
Finally, the impact of balancing the clusters in $k$-Means is paramount, as it brings in the critical information of the first dimension of the tile shape into the optimization algorithm.
In summary, our permutation algorithm variant improves the tile sparsity by \textbf{$\sim$4--1,000$\%$} for this network.

\subsection{Experimental Setup}\label{sec:exp-setup}

Our experiments involving the methods in Section~\ref{sec:methods} are done on a cluster of systems equipped with Tesla~A100 GPUs and Xeon~Gold 6258R CPUs.
We used PyTorch version 1.11.0 accelerated with CUDA~11.6.
For end-to-end evaluation using HElayers~\cite{helayers}, we used a 44-core Xeon~CPU E5-2699 @2.20~GHz machine with 750~GB memory. 
We observed that for our pruned networks, not all of the cores were fully utilized and performance saturated around 8 cores.
Thus, for a fair comparison, we limited the system to use 8 cores for all of our experiments.
We use a modified HElayers \cite{helayers} with the CKKS SEAL~\cite{sealcrypto} implementation targeting $128$-bit security, and the reported results are the average of $10$ runs. 

\mysec{Network Configurations and Model Hyperparameters} 
Table~\ref{tab:network_cfgs} shows details of the activation functions used in each network, the network architecture, and training hyperparameters for each of the networks and datasets in Section~\ref{sec:setup}.

\begin{table}[b]
\caption{Network configurations and hyperparameters. [HL=hidden layer]}
\label{tab:network_cfgs}
\resizebox{\columnwidth}{!}{
\begin{tabular}{|c||c|c|c|c|}
\hline
\textbf{Network} & \textbf{Activation Function}                                                                                  & \textbf{Configuration}                                                                                                                           & \textbf{Learning Rate Scheduler}                                                                                                                                        & \textbf{Learning Rate, Batch Size}                                                                                                               \\ \hline\hline
AE-Compressor    & 2\textsuperscript{nd} degree tanh~\cite{gottemukkula2019polynomial}                                                            & \begin{tabular}[c]{@{}c@{}}128 HL neurons (MNIST)\\ \{256,128,256\} HL neurons\\ (SVHN, CIFAR-10)\end{tabular}                                     & \begin{tabular}[c]{@{}c@{}}CosineAnnealingWarmRestarts\\ $T_0=5, \eta_{min}=10^{-4}$\end{tabular}                                                                       & \begin{tabular}[c]{@{}c@{}}$10^{-3}, 64$ (MNIST)\\ $10^{-3}, 128$ (CIFAR-10)\\ $10^{-4}, 64$ (SVHN)\end{tabular}                                 \\ \hline
AE-Denoiser      & 2\textsuperscript{nd} degree tanh~\cite{gottemukkula2019polynomial}                                                            & \begin{tabular}[c]{@{}c@{}}128 HL neurons (MNIST)\\ \{256,128,256\} HL neurons\\ (SVHN, CIFAR-10)\end{tabular}                                     & \begin{tabular}[c]{@{}c@{}}CosineAnnealingWarmRestarts\\ $T_0=5, \eta_{min}=10^{-4}$\end{tabular}                                                                       & \begin{tabular}[c]{@{}c@{}}$10^{-4}, 32$ (MNIST)\\ $10^{-4}, 64$ (SVHN, CIFAR-10)\end{tabular}                                                   \\ \hline
MLP-Classifier   & 2\textsuperscript{nd} degree tanh~\cite{gottemukkula2019polynomial}                                                            & \begin{tabular}[c]{@{}c@{}}128 HL neurons (MNIST)\\ \{256,128\} HL neurons\\ (SVHN, CIFAR-10)\end{tabular}                                         & \begin{tabular}[c]{@{}c@{}}CosineAnnealingWarmRestarts\\ $T_0=5, \eta_{min}=10^{-4}$\end{tabular}                                                                       & \begin{tabular}[c]{@{}c@{}}$10^{-3}, 64$ (MNIST)\\ $10^{-3}, 128$ (CIFAR-10)\\ $10^{-4}, 64$ (SVHN)\end{tabular}                                 \\ \hline
CNN-Classifier   & \begin{tabular}[c]{@{}c@{}}Square\\ (MNIST, SVHN, CIFAR-10)\\ Trained polynomials~\cite{baruch2022methodology}\\ (COVIDx CT-2A)\end{tabular} & \begin{tabular}[c]{@{}c@{}}32 HL input channels (MNIST)\\ \{32,64\} HL input channels\\ (SVHN, CIFAR-10)\\ See~\cite{baruch2022methodology} (COVIDx CT-2A)\end{tabular} & \begin{tabular}[c]{@{}c@{}}CosineAnnealingWarmRestarts\\ $T_0=5, \eta_{min}=10^{-4}$\\ (MNIST, SVHN, CIFAR-10)\\ ExponentialLR, $\gamma=0.99$\\ (COVIDx CT-2A)\end{tabular} & \begin{tabular}[c]{@{}c@{}}$10^{-3}, 64$ (MNIST)\\ $10^{-3}, 128$ (CIFAR-10)\\ $10^{-4}, 64$ (SVHN)\\ $10^{-6}, 128$ (COVIDx CT-2A)\end{tabular} \\ \hline
\end{tabular}
}
\end{table}

\begin{figure}[t]
     \centering
     \includegraphics[width=.7\linewidth]{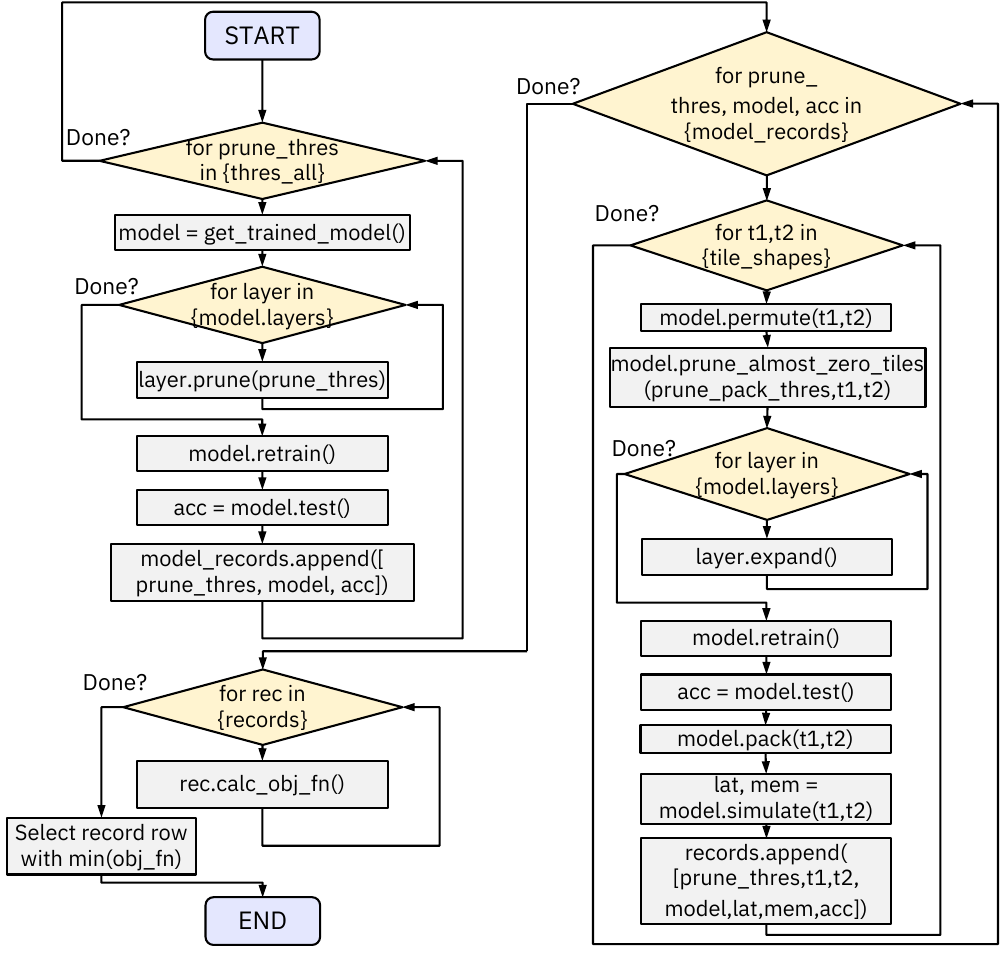}
     \caption{Flowchart illustrating the P4E strategy with local pruning scope for the prune step and global scope for prune\textsuperscript{pack}. This is executed as a pre-deployment step on the entity that has access to the plaintext network parameters.}
     \label{fig:flowchart}
\end{figure}

\mysec{Enhancing HElayers} 
We integrate \FRAME with HeLayers~\cite{helayers} and show a flowchart for P4E in Fig.~\ref{fig:flowchart}. 
This can be extended as needed to our other strategies.
We illustrate a grid search approach in the flowchart with \textit{\{thres\_all\}} and \textit{\{tile\_shapes\}} being a pre-selected set of values.
The output of the framework is a deployment-ready model that meets an objective function based on accuracy, latency, throughput, and memory.
A local search strategy may be considered to navigate the state space with fewer training and permutation rounds, which are the most time-consuming segments of this flow.
The zero-tile--skipping logic is programmed into HeLayers by associating a \texttt{zero\_flag} bit with each ciphertext, which is set during the HE $Enc$ procedure (Section~\ref{sec:he}).


%
%
%
\bibliographystyle{splncs04}
\bibliography{main}
%




\end{document}